\documentstyle[12pt,fullpage]{article}

\newcommand{\uncited}[1]{}

\date{}

\input{psfig.sty}

\newcommand{\fig}[4]
{\begin{figure}[htbp]
 \centerline{ \psfig{file=#1,height=#4}}
 \caption{#3}
 \label{#2}
 \end{figure}
}

\newcommand{\eps}{\varepsilon}

\newtheorem{theorem}{Theorem}
\newtheorem{lemma}[theorem]{Lemma}
\newtheorem{corollary}[theorem]{Corollary}

\newtheorem{conj}[theorem]{Conjecture}
\newtheorem{definition}{Definition}
\newtheorem{definitions}[definition]{Definitions}

\def\myendproof{\hfill{\vbox{\hrule\hbox{%
   \vrule height1.3ex\hskip0.8ex\vrule}\hrule}}}

\newcommand\real{{\rm I\kern-0.2em\rm R}}
\newcommand\itreal{{\it I\kern-0.35em\it R}}

\newenvironment{proof}{

\noindent{\bf Proof:}\ }{
\hfill \myendproof

}

\newenvironment{tabAlgorithm}[1]{
\setcounter{algorithmLine}{1}
\samepage
\begin{tabbing}
999\=\kill
#1
}{
\end{tabbing}
}
\newcounter{algorithmLine}
\newcommand{\algline}{\\\thealgorithmLine\hfil\>\stepcounter{algorithmLine}}
\newcommand{\algnono}{\\ \>}

\title{A Network-Flow Technique
  for Finding Low-Weight Bounded-Degree Spanning Trees\footnote{A preliminary 
version of this paper appeared in the
Proceedings of the 5th International Integer Programming and
Combinatorial Optimization Conference (IPCO), June 1996, pages 105-117.}}
\author{
    S\'andor P.~Fekete
  \thanks{Center for Parallel Computing, Universit\"at zu K\"oln,
          D-50923 K\"oln, Germany. E-Mail: {\tt sandor@zpr.uni-koeln.de.}}
   \and  Samir Khuller
  \thanks{Computer Science Department and Institute for Advanced Computer
    Studies, University of Maryland, College Park, MD~20742.
    Research supported by NSF Research Initiation Award CCR-9307462 and
    an NSF CAREER Award CCR-9501355.
    E-mail~: {\tt samir@cs.umd.edu}.}
  \and Monika Klemmstein
  \thanks{Center for Parallel Computing, Universit\"at zu K\"oln,
          D-50923 K\"oln, Germany. E-Mail: {\tt mklemmst@zpr.uni-koeln.de.}}
  \and Balaji Raghavachari
  \thanks{Department of Computer Science, The University of Texas at Dallas,
    Box 830688, Richardson, TX 75083-0688.
    Research supported in part by NSF Research Initiation Award CCR-9409625.
    E-mail : {\tt rbk@utdallas.edu}.}
  \and Neal Young
  \thanks{Dept. of Computer Science, Dartmouth College, Hanover NH 03755-3510. 
    Part of this research was done while at
    School of ORIE, Cornell University, Ithaca NY 14853
    and supported by \'Eva Tardos' NSF PYI grant DDM-9157199.
    E-mail : {\tt ney@cs.dartmouth.edu}.}
}

\begin{document}

\maketitle

\newcommand{\todo}[1]{{\bf [#1]}\marginpar{**}}

\newcommand{\mathfn}[1]{{\mathop{\rm #1}\nolimits }}
\newcommand{\goal}{\mathfn{d}} 
\newcommand{\degree}{\mathfn{deg}} 

\begin{abstract}
  Given a graph with edge weights satisfying the triangle inequality,
  and a degree bound for each vertex,
  the problem of computing a low weight spanning tree
  such that the degree of each vertex is at most its specified bound
  is considered.
  In particular, modifying a given spanning tree $T$
  using {\em adoptions} to meet the degree constraints is considered.
  A novel network-flow based algorithm for finding
  a good sequence of adoptions is introduced.
  The method yields a better performance guarantee
  than any previous algorithm.
  If the degree constraint $\goal(v)$ for each $v$ is at least $2$,
  the algorithm is guaranteed to find a tree whose weight is at most
  the weight of the given tree times
  $2 - \min\Big\{\frac{\goal(v)-2}{\degree_T(v)-2} : \degree_{T}(v)>2\Big\},$
  where $\degree_T(v)$ is the initial degree of $v$.
  Equally importantly,
  it takes this approach to the limit in the following sense:
  if any performance guarantee that is solely
  a function of the topology and edge weights of a given tree
  holds for {\em any} algorithm at all, then it also holds for the given algorithm.
  Examples are provided in which no lighter tree
  meeting the degree constraint exists.
  Linear-time algorithms are provided with the same worst-case
  performance guarantee.

  Choosing $T$ to be a minimum spanning tree yields approximation
  algorithms with factors less than 2
   for the general problem on geometric graphs with
  distances induced by various $L_p$ norms.
  Finally, examples of Euclidean graphs are provided in which
  the ratio of the lengths of an optimal Traveling Salesman path
  and a minimum spanning tree can be arbitrarily close to~2.
\end{abstract}


\section{Introduction}
Given a complete graph with edge weights satisfying the triangle inequality,
and a degree bound for each vertex,
we consider the problem of computing a low-weight spanning tree
in which the degree of each vertex is at most its given bound.
In general, it is NP-hard to find such a tree.
There are various practical motivations: the problem arises in the
context of VLSI layout and  
network design~\cite{Gv,MSh,St}
(such as in the Bellcore software {\em FIBER OPTIONS}, used for
designing survivable optimal fiber networks).
The special case of only one vertex with a degree-constraint
has been examined~\cite{Ga,GT,GlKl}; 
a polynomial time algorithm for the case of a fixed number of
nodes with a constrained degree was given by Brezovec et al.~\cite{BCG}. 
Computational results for
some heuristics for the general problem are presented
in~\cite{NH,SaVo,Vo}.
Papadimitriou and Vazirani~\cite{PV} raised the problem of finding the
complexity of computing a minimum-weight degree-4 spanning tree of
points in the plane.  Some geometric aspects are
considered in~\cite{KRY,MS,RS}.

In this paper, we consider modifying a given spanning tree $T$,
to meet the degree constraints without increasing its weight considerably.
We introduce a novel network-flow based algorithm
that does this optimally in the following sense:
if for some algorithm a worst-case performance guarantee
can be proved that is solely
a function of the topology and edge weights of $T$,
then that performance guarantee also holds for our algorithm.
We prove this by showing that our algorithm
finds the {\em optimal} solution for graphs in which
the weight of each edge $(u,v)$
equals the cost of the $u\leadsto v$ path in $T$.

We also show the following more concrete performance guarantee:
If the degree constraint $\goal(v)$ for each $v$ is at least $2$,
our algorithm finds a tree whose weight is at most
the weight of $T$ times
$$2 - \min\Big\{\frac{\goal(v)-2}{\degree_T(v)-2} : \degree_{T}(v)>2\Big\},$$
where $\degree_T(v)$ is the initial degree of $v$.
For instance, the degree of each vertex $v$ can be reduced by
nearly half, to $1+\lceil\degree_T(v)/2\rceil$, without increasing
the weight of the tree by more than $50 \%$.
(For comparison,
note that a factor of 2 is straightforward with standard shortcutting
techniques.)
We also describe {\em linear-time} algorithms that achieve this ratio.

This performance guarantee is optimal in the sense that for any $D\ge d \ge 2$,
if $T$ is a complete rooted $(D-1)$-ary tree with unit edge weights
and the edge weights in $G$ are those induced by paths in $T$,
then the weight of any spanning tree with maximum degree $d$ is at least
the weight of $T$ times $2- \frac{d-2}{D-2} - o(1)$.

The restriction $\goal(v) \ge 2$ is necessary to obtain constant
performance bounds.
Consider the case when $T$ is a simple path of unit weight edges, with
the remaining edge weights again induced by $T$.
Any spanning tree in which all but one vertex has degree one
is heavier than $T$ by a factor of $\Omega(n)$, the number of vertices in $T$.

For many metric spaces, graphs induced by points in the space
have minimum spanning trees of bounded maximum degree.
In such cases our algorithms can be used to find
spanning trees of even smaller degree
with weight bounded by a factor strictly smaller than 2 times
the weight of a minimum spanning tree (MST).
For example, in the $L_1$ metric,
a degree-4 MST can be found~\cite{RS},
so that we can find a degree-3 tree with weight at most $1.5$
times the weight of an MST.
We discuss similar results for the $L_1$, $L_2$, and $L_\infty$ norms.
For some of these norms, this improves the best current performance guarantees.

Finally, we disprove the following conjecture of \cite{KRY2}:
``In Euclidean graphs, perhaps a Traveling Salesman path
of weight at most $(2-\eps)$ times the minimum spanning-tree
weight always exists...''

Our algorithms modify the given tree by performing a sequence of {\em
adoptions}.
Our polynomial-time algorithm performs an optimal sequence of adoptions.
Adoptions have been previously used
to obtain bounded-degree trees in weighted graphs~\cite{KRY,RMRRH,Sal}.
The main contributions of this paper are a careful analysis
of the power of adoptions and a network-flow technique
for selecting an optimal sequence of adoptions.
The method yields a stronger performance guarantee
and may yield better results in practice.
The analysis of adoptions shows that different
techniques will be necessary if better bounds are to be obtained.


In the full version of their paper, Ravi et al.~\cite[Thm.~1.9]{RMRRH}
(if slightly generalized and improved\footnote{To obtain the improved bound
one has to change the proof slightly by
upper bounding $c(v_1v_2) - c(vv_2)$ by $c(v v_1)$ and not $c(v v_2)$
as is done in \cite{RMRRH}.})
gave an algorithm with a performance guarantee of
\[ 2- \min\left\{\frac{\goal(v)-2}{\degree_T(v)-1}
        : v\in V, \degree_{T}(v)>2\right\}
\]
provided each $\goal(v) \ge 3$.
The performance guarantee of our algorithm is better.

In Euclidean graphs (induced by points in $\real^d$),
minimum spanning trees are known to have bounded degree.
For such graphs, Khuller, Raghavachari and Young~\cite{KRY}
gave a linear-time algorithm to find a degree-3 spanning tree
of weight at most $5/3$ times the weight of a minimum spanning tree.
For points in the plane, the performance guarantee of their algorithm
improves to $1.5$; if the tree is allowed to have degree four,
the ratio improves further to $1.25$.

In unweighted graphs, F\"urer and Raghavachari \cite{FR}
gave a polynomial-time algorithm to find a spanning tree
of maximum degree exceeding the minimum possible by at most one.
In arbitrary weighted graphs, Fischer \cite{Fi} showed
that a minimum spanning tree
with maximum degree $O(\delta^*+ \log n)$ can be computed in polynomial time,
where $\delta^*$ is the minimum maximum degree
of any minimum spanning tree.  He also provided an algorithm
that finds a minimum spanning tree with degree
$k(\delta^*+1)$ where $k$ is the number of distinct edge weights.

\section{Adoption}
\begin{figure}[ht]
  \begin{center}
    \input{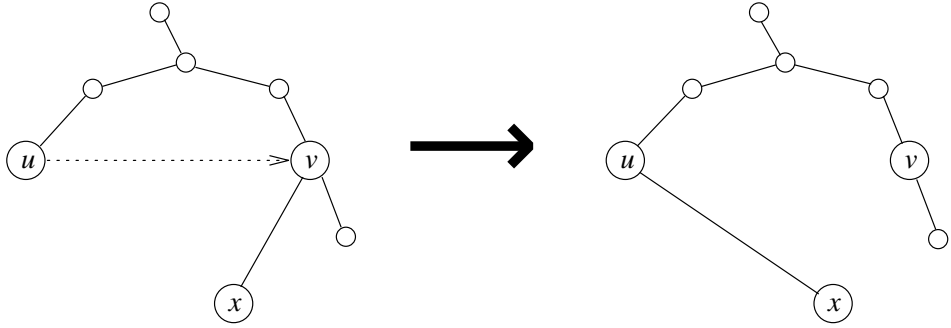}
  \end{center}
  \caption{Vertex $u$ {\em adopts} a neighbor of $v$}
  \label{adopt}
\end{figure}

Fix the graph $G=(V,V\times V)$
and the edge weights $w:V\times V\rightarrow \real$.
The algorithm starts with a given tree $T$
and modifies it by performing a sequence of {\em adoptions}.
The adoption operation (illustrated in Figure~\ref{adopt}) is as follows:

\begin{tabAlgorithm}{$\mbox{\sc Adopt}(u,v)$}
  \algnono {\bf Precondition:}
  Vertex $v$ has degree at least two in the current tree.
  \algline Choose a neighbor $x$ of $v$ in the current tree
  other than the neighbor on the 
\algnono current $u\leadsto v$ path.
  \algline Modify the current tree by replacing edge $(v,x)$ by $(u,x)$.
\end{tabAlgorithm}

The effect of $\mbox{\sc Adopt}(u,v)$ is that $u$ adopts a neighbor of
$v$.  For the purpose of the algorithm, it makes no difference which
neighbor of $v$ is adopted, since the differences in cost are hidden
by assigning $w(u,v)$ as the cost of the adoption as shown below.
$\mbox{\sc Adopt}(u,v)$ decreases the degree of $v$ by one,
at the expense of increasing the degree of $u$ by one
and increasing the weight of the tree by $w(x,u) - w(x,v) \le w(u,v)$.


\noindent
{\bf Comment:} In practice, we would wish to optimize the choice of
neighbor of $v$ we choose to be adopted, as this changes the weight
of the tree found. 


\subsection{The Adoption Network}
\begin{definitions}
  The {\em deficit}\/ of vertex $v$ with respect to $T$
  is $\degree_T(v)-\goal(v)$.
  Starting with a given tree, consider a sequence of adoptions
  $\mbox{\sc Adopt}(u_1,v_1),\mbox{\sc Adopt}(u_2,v_2),\ldots$
  \begin{itemize}
  \item The sequence is {\em legal}\/ if the precondition
    for each adoption is met.

  \item A sequence is {\em feasible}\/ if, for each vertex,
    the decrease in its degree, i.e., its old degree minus its new degree,
    is at least its deficit.

  \item The {\em cost}\/ of the sequence is $\sum_i w(u_i,v_i)$.

  \end{itemize}
\end{definitions}

The legal, feasible adoption sequences
are precisely those that yield a tree
meeting the degree constraints.
The cost of a sequence is an upper bound
on the resulting increase in the weight of the tree.
Our goal is to find a feasible legal sequence of minimum cost.
For brevity, we call such a sequence a minimum-cost sequence.

The problem reduces to a minimum-cost flow problem \cite{AMO} in a flow network
that we call the {\em adoption network}\/ for $T$.
The adoption network is defined as follows.
Starting with $G$, replace each edge $(u,v)$
by two directed edges $(u,v)$ and $(v,u)$,
each with cost $w(u,v)$ and infinite capacity.
Assign each vertex a demand equal to its deficit.

A {\em flow}\/ is an assignment of a real value
(called the flow on the edge) to each edge of the network.
The flow satisfies the following property, known as {\em skew
symmetry}~\cite{AMO}: $f(u,v) = -f(v,u)$ for any edge $(u,v)$.
For each vertex $v$, the {\em surplus}\/ at $v$
is the net flow assigned to incoming edges
minus the net flow assigned to outgoing edges.
Only edges with positive flow are considered in computing the surplus
flow into a vertex.
A flow is {\em legal}\/ if the surplus at each vertex
is at most one less than its degree.
A flow is {\em feasible}\/ if the surplus at each vertex
is at least its demand.
The {\em cost}\/ of the flow is the sum, over all edges,
of the cost of the edge times the flow on the edge.

Since the demands are integers,
there exists an integer-valued minimum-cost feasible flow~\cite{AMO}.
Assuming that each degree constraint is at least 1,
there exists such a flow that is also legal.
For brevity, we call such a flow a minimum-cost flow.

\begin{lemma} \label{correspondence-lemma}
The following statements are true:
\begin{enumerate}
  \item
  The adoption sequences correspond to integer-valued flows.
  The correspondence preserves legality, feasibility, and cost.
  \item
  The integer-valued flows correspond to adoption sequences.
  The correspondence preserves legality and feasibility;
  it does not increase cost.
\end{enumerate}
\end{lemma}
\begin{proof}
  Given a sequence of adoptions,
  the corresponding flow $f$ assigns a flow to each edge $(u,v)$
  equal to the number of times $u$ adopts a neighbor of $v$.
  It can be verified that this correspondence
  preserves legality, feasibility, and cost.

  Conversely, given an integer-valued flow $f$,
  modify it if necessary (by canceling flow around cycles)
  so that the set of edges with positive flow is acyclic.
  This does not increase the cost.
  Next, order the vertices so that, for each directed edge $(u,v)$
  with positive flow, $u$ precedes $v$ in the order.
  Consider the vertices in reverse order.
  For each vertex $u$, for each edge $(u,v)$ with positive flow,
  have $u$ adopt $f(u,v)$ neighbors of $v$.
Note that when an edge directed out of a vertex $v$ is processed by
{\sc Adopt}, $v$'s degree increases by one.  Similarly when an edge
directed into $v$ is processed, $v$'s degree decreases by one.
The order imposed above processes all outgoing edges (in the flow) of
a vertex before it processes any of its incoming edges.  Hence during
the course of the algorithm, $v$'s degree initially increases when its
outgoing edges are processed and subsequently decreases as its
incoming edges are processed.  Therefore, when {\sc Adopt} processes
an edge $(u,v)$, the precondition imposed by the procedure that $v$'s
degree be at least 2 always holds.
  It can be verified that the above sequence of adoptions
  preserves legality and feasibility
  and does not increase cost.
\end{proof}

\section{Polynomial-Time Algorithm}
An acyclic, integer, minimum-cost flow can be found in polynomial time~\cite{AMO}.
The corresponding legal, feasible adoption sequence
can be performed in polynomial time
as described in the proof
of the second part of Lemma~\ref{correspondence-lemma}.
This gives a polynomial-time algorithm.

\subsection{Optimality in Tree-Induced Metrics}
The following lemma shows that this algorithm is optimal
among algorithms that examine only the weights of edges
of the given tree.

\begin{lemma} \label{tree-optimal-lemma}
Given a weighted graph $G=(V,E)$ and a spanning tree $T$
such that the weight of each edge in $G$ equals
the weight of the corresponding path in $T$,
a minimum-cost sequence of adoptions yields an optimal tree.
\end{lemma}
\begin{proof}
Fix an optimal tree.
Note that the degree of $v$ in the optimal tree is at most
$\goal(v)$;  let it be ${\rm d}^*(v)$.
For each subset $S$ of vertices, let $\degree_T(S)$ and ${\rm d}^*(S)$
denote the sum of the degrees of vertices in $S$ in $T$ and in the optimal
tree, respectively.
Define a flow on the edges of $T$ as follows:
for each edge $(u,v)$ in $T$, let $f(u,v) = {\rm d}^*(S_u)-\degree_T(S_u)$,
where $S_u$ is the set of vertices that are reachable from $u$ using
edges in $T$ other than $(u,v)$.
Note that $f(u,v) = -f(v,u)$.  Intuitively, a negative flow of $x$
units from $u$ to $v$ means that $x$ units of flow go from $v$ to $u$.
This is known as the {\em skew symmetry} property of flows~\cite{AMO}.
Inductively it can be shown that for each vertex $v$,
the net flow into it is $\degree_T(v) - {\rm d}^*(v)$,
so that the adoption sequence determined by the flow $f$
achieves a tree with the same degrees as the optimal tree.

We will show that the cost of the flow, and therefore the cost
of the adoption sequence, is at most the difference in the weights
of the two trees.  This implies that the tree obtained by the adoption
sequence is also an optimal tree.

To bound the cost of the flow, we claim that the flow is ``necessary''
in the following sense:
for each edge $(u,v)$ in $T$, at least $f(u,v)+1$ edges in the optimal tree
have one endpoint in $S_u$ and the other in $V-S_u$.
To prove this,
let $c$ be the number of edges in the optimal tree crossing the cut $(S_u,V-S_u)$.
Note that $\degree_T(S_u) = 2(|S_u|-1) + 1$.
Since the optimal tree is acyclic,
the number of edges in the optimal tree
with both endpoints in $S_u$ is at most $|S_u|-1$.
Thus ${\rm d}^*(S_u) \le 2(|S_u|-1) + c = \degree_T(S_u)-1 + c$.
Rewriting gives $c \ge {\rm d}^*(S_u) - \degree_T(S_u) + 1 = f(u,v) + 1$.
This proves the claim.

To bound the cost of the flow, for each edge $(u,v)$,
charge $w(u,v)$ units to each edge in the optimal tree crossing
the cut $(S_u,V-S_u)$.
By the claim, at least the cost of the flow, plus the cost of $T$, is charged.
However, since the cost of each edge in the optimal tree
equals the weight of the corresponding path in $T$,
each edge in the optimal tree is charged at most its weight.
Thus, the total charge assigned to the edges
is bounded by the weight of the optimal tree.
\end{proof}

Note that given the exact degrees of the desired tree
(for instance, if the degree constraints sum to $2(|V|-1)$),
the optimal flow in Lemma~\ref{tree-optimal-lemma}
can be computed in linear time.

\subsection{Worst-Case Performance Guarantee}
The next theorem establishes a worst-case performance guarantee for
the algorithm in general graphs satisfying the triangle inequality.
\begin{theorem} \label{performance-theorem}
Given a graph $G=(V,E)$ with edge weights satisfying the triangle inequality,
a spanning tree $T$, and, for each vertex $v$, a degree constraint $\goal(v) \ge 2$,
the algorithm produces a tree whose weight is at most the weight of $T$ times
\begin{displaymath}
2- \min\left\{\frac{\goal(v)-2}{\degree_T(v)-2} : v\in V, \degree_{T}(v)>2\right\}.
\end{displaymath}
\end{theorem}
\begin{proof}
The increase in the cost of the tree
is at most the cost of the best sequence.
By Lemma~\ref{correspondence-lemma}, 
this is bounded by the cost of the minimum-cost flow.
We exhibit a fractional feasible, legal flow whose cost is
appropriately bounded.
The minimum-cost flow is guaranteed to be at least as good.

  Root the tree $T$ at an arbitrary vertex $r$.
  Push a uniform amount of flow along each edge towards the root
  as follows.  Let $p(v)$ be the parent of each non-root vertex $v$.
  For a constant $c$ to be determined later, define
  \begin{displaymath}
    f(u,v) = \cases{c & if $v = p(u)$ \cr 0 & otherwise.}
  \end{displaymath}
  The cost of the flow is $c$ times the weight of $T$.
  Let $v$ be any vertex.
  The surplus at $v$ is at least $c(\degree_T(v)-2)$.
  We choose $c$ just large enough so that the flow is feasible.

  There are three cases.
  If $\degree_T(v) = 1$,
  the deficit at $v$ will be satisfied provided $c \le 1$ and $\goal(v)\ge 2$.
  If $\degree_T(v) = 2$,
  the deficit at $v$ will be satisfied provided $\goal(v)\ge 2$.
  For $\degree_T(v) > 2$, the deficit will be satisfied provided
  \begin{displaymath}
    c \ge \frac{\degree_T(v)-\goal(v)}{\degree_T(v)-2}
    = 1-\frac{\goal(v)-2}{\degree_T(v)-2}.
  \end{displaymath}
  Thus, taking
  \begin{displaymath}
    c = 1- \min\left\{\frac{\goal(v)-2}{\degree_T(v)-2}
                         : v\in V, \degree_{T}(v)>2\right\}
  \end{displaymath}
  gives the result.
\end{proof}


\section{Optimality of Performance Guarantee}
In this section, we show that the worst-case performance guarantee established
in Theorem~\ref{performance-theorem} is the best 
obtainable.

\begin{lemma}
Consider an $n$-vertex weighted graph $G$ with a spanning tree $T$
such that the weight of each edge in $T$ is $1$ and
the weight of each remaining edge
is the weight of the corresponding path in $T$.
If $T$ corresponds to a complete rooted $(D-1)$-ary tree of depth $k$,
then the weight of any spanning tree with maximum degree $d$ is at least
the weight of $T$ times
\begin{displaymath}
2- \frac{d-2}{D-2} - o(1),
\end{displaymath}
where $o(1)$ tends to $0$ as $n$ grows. 
\end{lemma}
\begin{proof}
Fix any spanning tree $T'$ of maximum degree $d$.
Let $S_i$ denote the vertices at distance at most $i$ from the root in $T$.
The flow on the edges of $T$ corresponding to $T'$, as defined in
Lemma~\ref{tree-optimal-lemma}, can be generalized to arbitrary cuts
$(V-S,S)$ in the tree, and it can be shown that the flow crossing this
cut is at least $\degree_{T}(S) - \degree_{T'}(S)$.
For any $i<k$, the cut $(V-S_i,S_i)$ is crossed by at least
$|S_i|(D-d)-1$ units of flow.
Thus the total cost of the flow is at least $\sum_{i=0}^{k-1}(|S_i|(D-d)-1)$.
The cost of $T$ is $|S_k|-1$, which can be written as
$\sum_{i=0}^{k-1} |S_{i+1}|-|S_i|$.
It can be verified that $|S_{i+1}|-|S_i| = |S_i|(D-2) + 1$.
Hence the ratio of the cost of the flow to the cost of $T$ is at least
$$\frac{\sum_{i=0}^{k-1}(|S_i|(D-d)-1)}{\sum_{i=0}^{k-1}(|S_i|(D-2)+1)}.$$
On simplification, it can be seen that the ratio is
at least $(D-d)/(D-2) - o(1)$.
Since the ratio of the cost of $T'$ to the cost of $T$
is $1$ more than this, the result follows.
\end{proof}

Next we observe that the $\goal(v) \ge 2$ constraint 
is necessary to obtain any constant performance guarantee:

\begin{lemma}
Consider an $n$-vertex weighted graph $G$ with a spanning tree $T$
such that the weight of each edge in $T$ is $1$ and
the weight of each remaining edge
is the weight of the corresponding path in $T$.
If $T$ corresponds to a path of length $n$ with endpoint $r$,
then the weight of any spanning tree in which each vertex other than $r$
has degree $1$ is at least the weight of $T$ times $n/2$.
\end{lemma}
The proof is straightforward.


\section{Linear-Time Algorithms}
Note that to obtain the worst-case performance guarantee
a minimum-cost flow is not required.
It suffices to find a feasible integer flow
of cost bounded by the cost of the fractional flow $f$
defined in the proof of Theorem~\ref{performance-theorem}.
We describe two methods to find such a flow,
and to implement the corresponding sequence of adoptions,
in linear time.

\paragraph{Algorithm 1:}
The first algorithm exploits the special structure of the flow in the
proof of Theorem~\ref{performance-theorem} to construct an integral
flow without solving the flow problem to optimality.  Observe that the
flow along each edge of the tree is the same ($c$ units).  Since the
graph satisfies the triangle inequality, flow along a path of more
than a single edge can be replaced by a single edge that connects the
end vertices of the path without increasing the total cost of the
flow.  This ensures that all of the fractional flows are sent from the
sources to the destinations directly.  The following greedy rounding scheme finds
an integral flow that is no more expensive than the fractional flow.

Let $f$ be the fractional flow defined in Theorem~\ref{performance-theorem}.
Modify $f$ by repeatedly performing the following short-cutting step:
choose a maximal path in the set of edges with positive flow;
replace the ($c$ units of) flow on the path
by ($c$ units of) flow on the single new edge $(u,v)$,
where the path goes from $u$ to $v$.
Let $q(u)$ be the child of $v$ on the path.
Stop when all paths have been replaced by new edges.
This phase requires linear time,
because each step requires time
proportional to the number of edges short-cut.

In the resulting flow, the only edges with positive flow
are edges from leaves of the (rooted) tree $T$ to interior vertices.
Round the flow to an integer flow as follows.
Consider each vertex $v$ with positive deficit, say $D$.
Using a linear-time selection algorithm,
among the edges $(u,v)$ sending flow to $v$,
find the $D$ smallest-weighted edges.
Assign one unit of flow to each of these $D$ edges.
The resulting flow is integer-valued, feasible, legal, and has cost
bounded by the cost of $f$.
This phase requires linear time.

Assume that each vertex maintains a doubly linked list of its children.
Given a pointer to any vertex, we can obtain its sibling in constant
time. As adoptions are done, this list is maintained dynamically.
Perform the adoptions corresponding to the flow in any order:
for each edge $(u,v)$ with a unit of flow,
have $u$ adopt the right sibling of $q(u)$ (in the original tree $T$).
The tree remains connected because $\goal(v) \ge 2$,
so at least one child of $v$ is not adopted.

\paragraph{Algorithm 2:}
Consider the following restricted adoption network.
Root the tree $T$ as in the proof of Theorem~\ref{performance-theorem}.
Direct each edge $(u,v)$ of the tree towards the root.
(Non-tree edges are not used.)
Assign each edge a capacity of 1 and a cost equal to its weight.
Assign each vertex a demand equal to its deficit.

We show below that an integer-valued minimum-cost flow
in this network can be found in linear time.
Because the fractional flow
defined in the proof of Theorem~\ref{performance-theorem}
is a feasible legal flow in this network,
the minimum-cost flow that we find is at least as good.

Find the flow via dynamic programming.
For each vertex $v$,
consider the subnetwork corresponding to the subtree rooted at $v$.
Let $C_j(v)$ denote the minimum cost of a flow in this subnetwork
such that the surplus at $v$ exceeds its demand $D$ by $j$, for $j=0,1$.
Since the flow problem has been restricted to the tree, with a
capacity constraint of 1 on all the edges, there is no need to
consider flow surpluses greater than 1 at any vertex.
For each child $u$ of $v$,
let $\delta(u)$ denote $w(u,v)+C_1(u)-C_0(u)$ ---
the additional cost incurred for $v$ to obtain
a unit of flow along edge $(u,v)$.
Let $U_j$ denote the $D+j$ children with smallest $\delta(u)$, for $j=0,1$.
Then, for $j=0,1$,
$$C_j(v) = \sum_{u\in U_j} \delta(u) + \sum_u C_0(u).$$
Using this equation, compute the $C_j$'s bottom-up in linear time.
The cost of the minimum-cost flow in the restricted network
is given by $C_0(r)$, where $r$ is the root.
The flow itself is easily recovered in linear time.

To finish, shortcut the flow as in
the first phase of the previous algorithm
and perform the adoptions as in the last phase of that algorithm.

\section{Geometric Problems}
Our general result has several implications for cases of particular
distance functions where it is possible to give a priori bounds on the
maximum degree of an MST.
For the case of $L_2$ distances in the plane,
there always is an MST of maximum degree 5~\cite{MS};
for the case of $L_1$ or $L_{\infty}$ distances
there always exists a
MST of maximum degree $4$~\cite{MS,RS}.
Without using any specific structure of the involved distance functions,
we note as a corollary:

\begin{corollary}
\label{geo}
Let $T_{\min}$ be an MST and $T_k$ be a tree whose maximal degree is
at most $k$.
For $L_1$ or $L_{\infty}$ distances in $\real^2$, we get a degree-3
tree $T_3$ with
\begin{itemize}
\item $w(T_3)<\frac{3}{2}w(T_{\min})$.
\end{itemize}
\end{corollary}

\noindent
For the case of Euclidean distances in the plane, we get bounded degree
trees that satisfy
\begin{itemize}
\item $w(T_3)<\frac{5}{3} w(T_{\min})$
\item $w(T_4)<\frac{4}{3} w(T_{\min})$.
\end{itemize}
The latter two bounds are worse than those shown by
Khuller, Raghavachari and Young~\cite{KRY}
using the geometry of point
arrangements. (It was shown that $\frac{3}{2}$ and $\frac{5}{4}$
are upper bounds.) We conjecture that the following are
the optimal ratios:

\begin{conj}
\label{bounds}
For the case of Euclidean distances in the plane, we conjecture
that there exist bounded degree trees that satisfy
\begin{itemize}
\item $\frac{w(T_3)}{w(T_{\min})}\leq \frac{\sqrt{2}+3}{4} \approx 1.103\ldots$
\item $\frac{w(T_4)}{w(T_{\min})}\leq \frac{2\sin(\frac{\pi}{10})+4}{5}
                                        \approx 1.035\ldots$
\end{itemize}
For $L_1$ and $L_{\infty}$ distances in $\real^2$, we conjecture
\begin{itemize}
\item $\frac{w(T_3)}{w(T_{\min})}\leq\frac{5}{4}$
\end{itemize}
\end{conj}

The worst examples we know (matching the ratios
of Conjecture~\ref{bounds}) are shown in Figure~\ref{best}.
(Note that the example for $L_{\infty}$ metric is obtained
by rotating the arrangement in (c) by 45 degrees.)

\fig{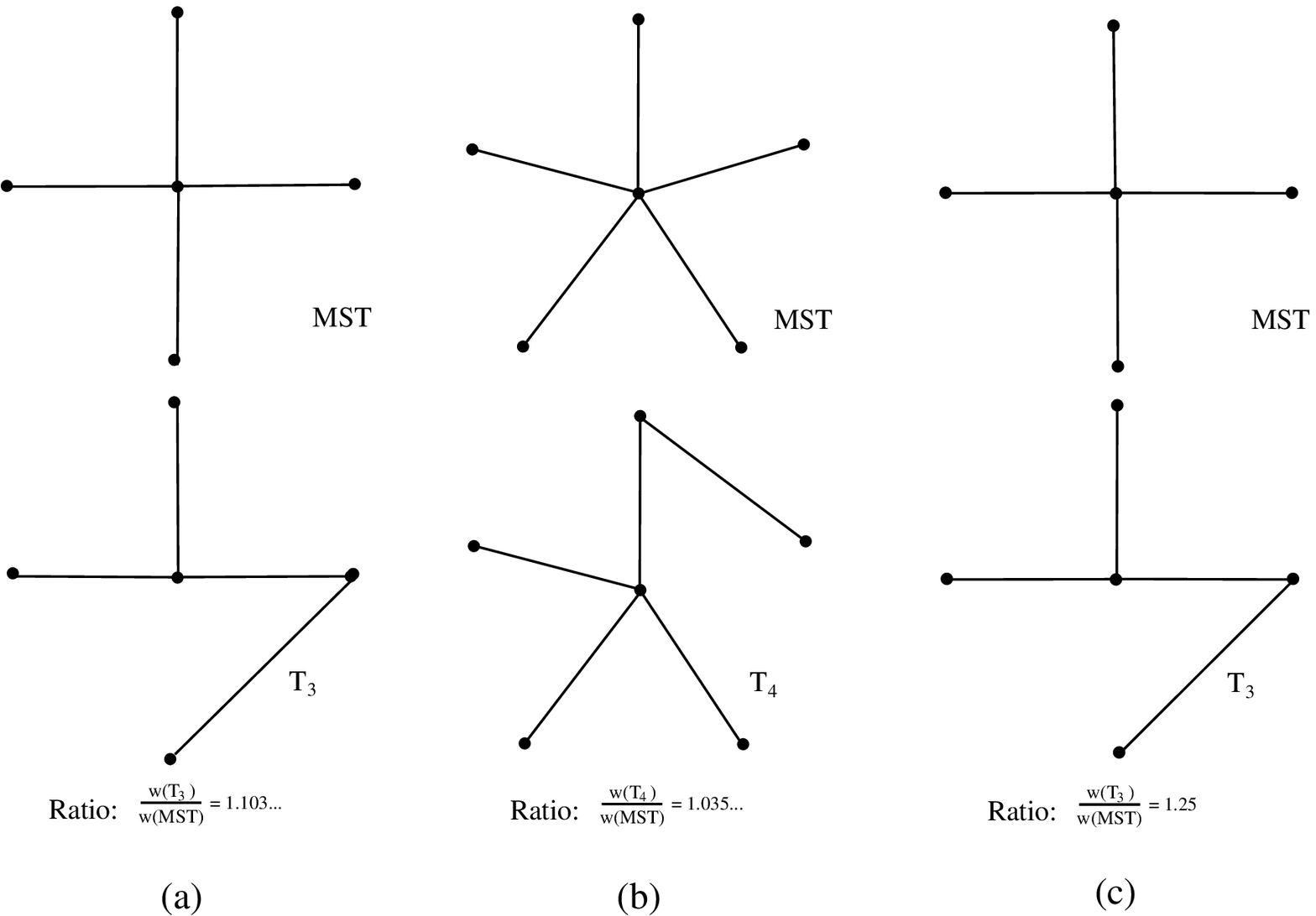}{best}{The worst known examples for:
(a) $\frac{w(T_3)}{w(T_{\min})}$, $L_2$ distances
\newline\hspace*{2.79in}(b) $\frac{w(T_4)}{w(T_{\min})}$, $L_2$ distances
\newline\hspace*{2.79in}(c) $\frac{w(T_3)}{w(T_{\min})}$, $L_1$
distances.}{2.5in}

\vspace{1ex}
\subsection{Geometric Hamiltonian Paths}
We conclude this paper by settling a question
raised in~\cite{KRY2}, in the negative:

{\em ``In Euclidean graphs, perhaps a Traveling Salesman path
of weight at most $(2-\eps)$ times the minimum spanning-tree
weight always exists and can be found in polynomial time.''}

\begin{theorem}
\label{T2}
For an arrangement of points in the plane with Euclidean
distances, the ratio $\frac{w(T_2)}{w(T_{\min})}$
can be arbitrarily close to 2.
\end{theorem}

\fig{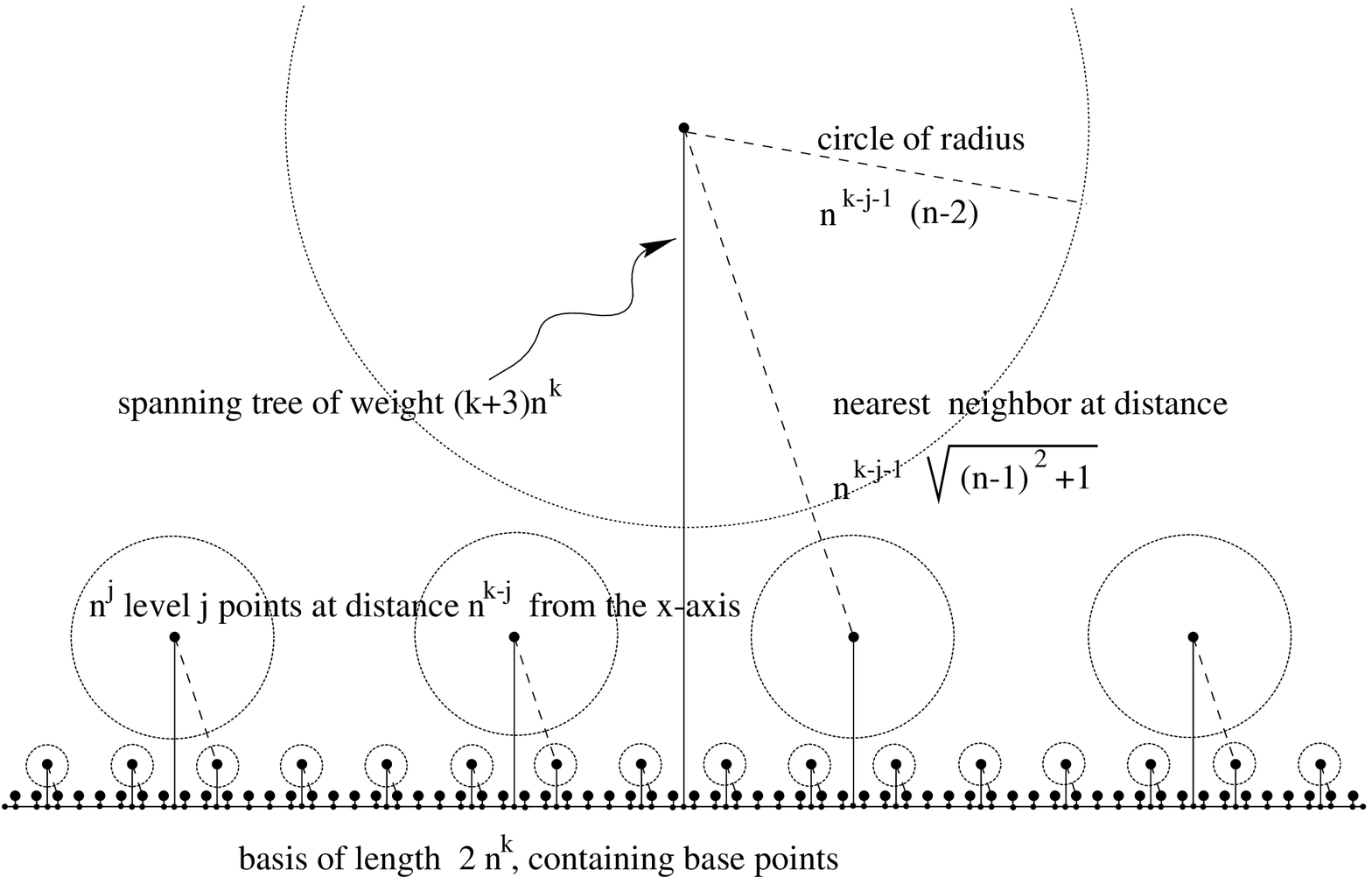}{T2.bw}{A class of examples showing 
$\frac{w(T_2)}{w({T_{\min}})}\rightarrow 2$}{2.0in}

\begin{proof}
Let $n$ and $k$ be sufficiently large.
Construct a point set as follows (see Figure~\ref{T2.bw}):

Take base points at $(0,0)$ and $(2n^k,0)$.

For $j=0,\ldots, k$, add points as follows:

\ \ For $i=1,\ldots, n^j$

\ \ \ \ Add {\em level} $j$ point at $((2i-1)n^{k-j},n^{k-j})$;

\ \ \ \ add {\em base} point at $((2i-1)n^{k-j},0)$.

The  points at level $j$, i.e., at height $n^{k-j}$, have nearest neighbors at
distance at least $n^{k-j-1}(n-1)$. To prove the lower bound,
we draw a circle centered at each point at level $j<k$. For the points
at level $j$, the radius of the circle is $(n^{k-j-1}(n-2))$.
The circles corresponding to two points do not intersect.
Since each point has degree two in a Hamilton cycle, twice the sum of the
radii of the circles gives us a lower bound on the length of the
Hamilton cycle. This can be computed as follows (observe that we can
always pick $n \geq 2k$).
\[ 2 \sum_{j=0}^{k-1} n^j (n^{k-j-1}(n-2)) = 2 k n^{k-1}(n-2)
\geq 2 (k-1)n^k .\]
Since no edge can have
length more than $2n^k$, we conclude that no Hamilton {\em path}
can have a weight smaller than $2 (k-1)n^k-2n^k = 2(k-2) n^k$.

It can be verified that there is a tree of weight $(k+3)n^k$ that spans
the points. Hence this is an upper bound on the weight of $T_{\min}$.
It follows that $\frac{w(T_2)}{w(T_{\min})}>\frac{2(k-2)}{k+3}$,
which can be arbitrarily close to 2, concluding the proof.
\end{proof}

The above class of examples establishes
the same lower bound for $L_1$ and $L_{\infty}$
distances.


\section*{Acknowledgements}
We thank Joe Mitchell for establishing the transatlantic
connection between the authors.
We thank Chandra Chekuri for asking us about degree 3 trees
in the $L_1$ metric.


\begin{thebibliography}{CDNS}

\bibitem{AMO}
R.~K.~Ahuja, T.~L.~Magnanti and J.~B.~Orlin.
\newblock {\em Network flows (theory, algorithms and applications)}.
\newblock Prentice Hall, Englewood Cliffs, NJ, 1993.

\uncited{
\bibitem{BGMRZ}
T.~Barrera, J.~Griffith, A.~McKee, G.~Robins, and T.~Zhang.
\newblock Toward a Steiner engine: enhanced serial and parallel
        implementations of the iterated 1-Steiner algorithm.
\newblock {\em Proc.\ Great Lakes Symp.\ VLSI}.
\newblock Kalamazoo, MI, March 1993, pp.\ 90--94.
}

\uncited{
\bibitem{B}
Bellcore.
\newblock {\em FIBER OPTIONS}.
Software for designing survivable optimal fiber networks.
Software package, 1988.
}


\bibitem{BCG}
C.~Brezovec, G.~Cornu\'ejols, F.~Glover.
\newblock A matroid algorithm and its application
to the efficient solution of two optimization
problems in graphs.
\newblock {\em Math.\ Programming} {\bf 42} (1988), pp.\ 471--487.

\bibitem{Fi}
T.~Fischer.
\newblock {\em Optimizing the degree of minimum weight spanning trees}.
\newblock Tech.\ Rep.\ 93-1338, Dept.\ of Computer Science,
        Cornell University, April 1993.

\bibitem{FR}
M.~F\"urer and B.~Raghavachari.
\newblock Approximating the minimum-degree Steiner tree to within one of
optimal.
\newblock  {\em J.\ Algorithms} {\bf 17} (1994), pp.\ 409--423.

\bibitem{Ga}
H.~N.~Gabow.
\newblock A good algorithm for smallest spanning trees with
    a degree constraint.
\newblock {\em Networks} {\bf 8} (1978), pp.\ 201--208.

\bibitem{GT}
H.~N.~Gabow and R.~E.~Tarjan.
\newblock Efficient algorithms for a family of matroid intersection problems.
\newblock {\em J.\ Algorithms} {\bf 5} (1984), pp.\ 80--131.

\bibitem{GJ}
M.~R.~Garey and D.~S.~Johnson.
\newblock {\em Computers and intractability: a guide to the
theory of NP-completeness}.
\newblock Freeman, San Francisco, CA, 1979.

\bibitem{Gv}
B.~Gavish.
\newblock Topological design of centralized
        computer networks --- formulations and algorithms.
\newblock {\em Networks} {\bf 12} (1982), pp.\ 355--377.

\uncited{
\bibitem{GP}
G.~Georgakopoulos and C.~H.~Papadimitriou
\newblock The 1-Steiner tree problem.
\newblock {\em J.~Algorithms}, {\bf 8} (1987), pp.\ 122--130.
}

\bibitem{GlKl}
F.~Glover, D.~Klingman.
\newblock Finding minimum spanning trees with a fixed number of links at
a node.
\newblock In: B.\ Roy (ed.), {\em Combinatorial Programming: Methods and
Applications}.
\newblock D.~Reidel Publishing Company, Dordrecht-Holland, 1975. pp.\ 191--201.

\uncited{
\bibitem{Go}
M.~C.~Goldstein.
\newblock Design of long-distance telecommunication networks
-- the Telepak problem.
\newblock {\em IEEE Trans.\ Circuit Theor.} CT-20 (1973), pp.\ 186--192.
}

\uncited{
\bibitem{Ha}
A.~C.~Harter.
\newblock {\em Three-dimensional circuit layout}.
\newblock Cambridge University Press, New York, 1991.
}

\uncited{
\bibitem{H}
J.~A.~Hoogeveen.
\newblock Analysis of Christofides' heuristic: some paths are
more difficult than cycles.
\newblock {\em Operations Research Letters}, {\bf 10}, 291--295, 1991.
}

\uncited{
\bibitem{KR}
A.~B.~Kahng and G.~Robins.
\newblock A new class of iterative Steiner tree heuristics
with good performance.
\newblock {\em IEEE Trans.\ Computer-Aided Design}, {\bf 11} (1992), 893--902.
}

\bibitem{KRY}
S.~Khuller, B.~Raghavachari and N.~Young.
\newblock Low degree spanning trees of small weight.
\newblock {\em SIAM J.\ Comput.} {\bf 25} (1996), pp.\ 355--368. 

\bibitem{KRY2}
S.~Khuller, B.~Raghavachari, N.~Young,
\newblock Balancing minimum spanning trees and shortest-path trees.
\newblock {\em Algorithmica} {\bf 14} (1995), pp.\ 305--321.

\uncited{
\bibitem{Le}
T.~Lengauer.
{\em Combinatorial algorithms for integrated circuit layout}.
B.~G.~Teubner, Stuttgart, 1990.  
}

\bibitem{MSh}
C.~L.~Monma, D.~Shallcross.
\newblock Methods for designing communication networks with certain
two-connected survivability constraints.
\newblock {\em Oper.\ Res.} {\bf 37} (1989), pp.~531--541.

\bibitem{MS}
C.~L.~Monma, S.~Suri,
\newblock
 Transitions in geometric minimum spanning trees.
\newblock
 {\em Discrete \& Computational Geometry} {\bf 8} (1992), pp.\ 265--293.

\bibitem{NH}
S.~C.~Narula and C.~A.~Ho.
\newblock Degree-constrained minimum spanning tree.
\newblock {\em Comput.\ Ops.\ Res.} {\bf 7} (1980), pp. 239--249.

\bibitem{PV}
C.~H.~Papadimitriou, U.~V.~Vazirani,
\newblock
 On two geometric problems related to the traveling salesman problem.
\newblock
 {\em J. Algorithms} {\bf 5} (1984), pp.\ 231--246.

\uncited{
\bibitem{PL}
B.~T.~Preas and M.~J.~Lorenzetti.
{\em Physical design automation of VLSI systems}.
Benjamin/Cummings, Menlo Park, CA, 1988.
}

\uncited{
\bibitem{Pr}
R.~C.~Prim.
\newblock Shortest connection networks and some generalizations.
\newblock {\em Bell Systems Tech.~J.} {\bf 36} (1957), pp.\ 1389--1401.
}

\bibitem{RMRRH}
R.\ Ravi, M.\ V.\ Marathe, S.\ S.\ Ravi, D.\ J.\ Rosen\-krantz
        and H.\ B.\ Hunt\ III.
\newblock
Many birds with one stone: multi-objective approximation algorithms.
\newblock Manuscript.
\newblock A preliminary version appeared in
{\em Proc.~25th Annual ACM Symp. on the Theory of Computing},
pp.~438--447, May 1993.

\uncited{
\bibitem{RMRRH2}
R.\ Ravi, M.\ V.\ Marathe, S.\ S.\ Ravi, D.\ J.\ Rosen\-krantz
        and H.\ B.\ Hunt\ III.
\newblock
Many birds with one stone: multi-objective approximation algorithms.
\newblock
Manuscript (full version of 1993 STOC paper).
}

\bibitem{RS}
G.~Robins and J.~S.~Salowe.
\newblock Low-degree minimum spanning trees.
\newblock {\em Discrete and  Computational Geometry} {\bf 14} (1995),
                pp.\ 151--166.

\bibitem{Sal}
J.~S.~Salowe.
\newblock Euclidean spanner graphs with degree four.
\newblock {\em Discrete Appl. Math.} {\bf 54} (1994), pp.\ 55--66.

\bibitem{SaVo}
M.~Savelsbergh and A.~Volgenant.
\newblock Edge exchanges in the degree-constrained minimum
 spanning tree problem.
\newblock {\em Comput.\ Ops.\ Res.} {\bf 12} (1985), pp.\ 341--348.

\bibitem{St}
M.~Stoer.
\newblock {\em Design of survivable networks}.
\newblock Lecture Notes on Mathematics \# 1531.
\newblock Springer, Heidelberg, 1992.

\bibitem{Vo}
A.~Volgenant.
\newblock A Lagrangean approach to the degree-constrained minimum
 spanning tree problem.
\newblock {\em Europ.\ J.\ Ops.\ Res.} {\bf 39} (1989), pp.\ 325--331.

\uncited{
\bibitem{Y}
Y.~Yamamoto.
\newblock The Held-Karp algorithm and degree-constrained
minimum 1-trees.
newblock {\em Mathematical Programming \/} {\bf 15} (1978), pp.\ 228--231.
}

\end{thebibliography}
\end{document}